\def\year{2017}\relax
\documentclass[letterpaper]{article} 
\usepackage{aaai17}  
\usepackage{times}  
\usepackage{helvet}  
\usepackage{courier}  
\usepackage{url}  
\usepackage{graphicx}  
\begin{document}
%
\title{Detection of social signals for recognizing engagement in human-robot interaction}
\author{Divesh Lala, Koji Inoue, Pierrick Milhorat and Tatsuya Kawahara\\
Graduate School of Informatics, Sakyo-ku\\
Kyoto University\\
Kyoto, Japan\\
}
\maketitle
\begin{abstract}
Detection of engagement during a conversation is an important function of human-robot interaction. The level of user engagement can influence the dialogue strategy of the robot. Our motivation in this work is to detect several behaviors which will be used as social signal inputs for a real-time engagement recognition model. These behaviors are nodding, laughter, verbal backchannels and eye gaze. We describe models of these behaviors which have been learned from a large corpus of human-robot interactions with the android robot ERICA. Input data to the models comes from a Kinect sensor and a microphone array. Using our engagement recognition model, we can achieve reasonable performance using the inputs from automatic social signal detection, compared to using manual annotation as input.
\end{abstract}

\section{Introduction}
Human conversation makes use of a range of non-verbal social signals, which provide additional information about the internal state of the parties involved. Conversational robots should be able to recognize such signals. By doing this, they can control their behavior and dialogue to provide more natural communication with humans.

We assume that some social signals are used to indicate the engagement of a user. Engagement has been described as the process of establishing, maintaining an interaction \cite{Sidner2005} or, more concretely, how interested and attentive they are towards a conversation \cite{Yu2004}. One major goal is to keep the user engaged in a conversation. Therefore, if the robot is able to detect a change in the engagement level of the user, we can formulate a dialogue strategy to make or keep them engaged, thereby improving the user experience. For example, the robot may choose to continue or discontinue the current topic of conversation depending on how engaged it perceives the user to be. This type of scenario motivates the work in this paper.

\begin{figure}
	\centering
		\includegraphics{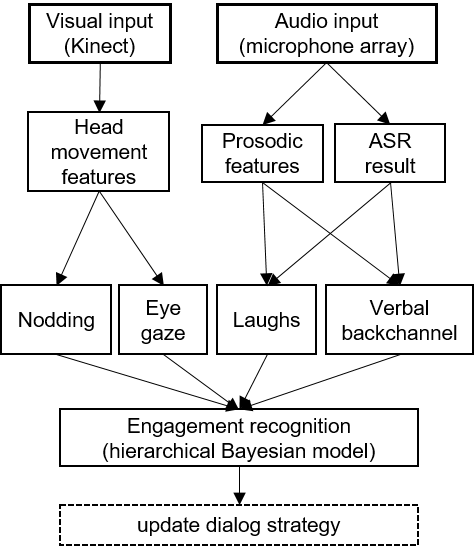}
	\caption{An overview of the architecture of the engagement recognition system.}
		\label{engagementmodel}
\end{figure}

While engagement cannot be measured directly, we can infer it by observing a number of social signals. In this paper, we discuss four of these - nodding, laughter, verbal backchannels, and eye gaze. These have been identified in other works as being indicative of the engagement level of the user \cite{Rich2010,Xu2013,Oertel2015}. The general architecture of our engagement system is shown in Figure \ref{engagementmodel}. We use Kinect and a microphone array as sensors in our system. Our objective in this paper is to implement models which detect the social signals in real-time so that they can be used as inputs to the engagement recognition model.

The robot which we use for this work is ERICA, a specially designed android who is a young Japanese woman \cite{Glas2016}. Our long-term goal is for ERICA to interact with humans in the real world. She has a human-like appearance and over 20 motors in her face which generate life-like facial movements for speech and emotion expression. ERICA can be tele-operated by another human, but we intend for her to be completely autonomous with the ability to recognize and respond appropriately to natural speech. Currently ERICA is integrated with a speech recognition system and a module to track the subject's body movements. We use these sensors to train our social signal recognition models.

We expect that ERICA's realism will elicit humans to interact with her as if she is a real human. In previous studies we have developed ERICA's dialogue system \cite{Lala2017,Milhorat2017}, but have not considered the state of the user. We are now integrating engagement recognition into ERICA's architecture. Our approach in this work is to create a large corpus and annotate social signals related to engagement during interactions with her. The signals occur within a human-robot conversation but are natural behaviors seen in the real world. We then use this data to train social signal detection models for the engagement recognizer. We then compare the performance of the engagement recognition model when using the detection models with manual annotation as input.


\section{Related Work}
Tracking user engagement for agent interaction has been an active area of research \cite{Rich2010,Forbes2012,Yu2016,Yu2017}. The general approach is to discover what types of social signals are related to engagement and then create a model which can measure the engagement level of the user. This is often a binary or categorical classification \cite{Bednarik2012}. 

The ability to recognize engagement in the user can influence the conversation. Previous works have found that engagement is related to turn-taking \cite{Xu2013,Inoue2016}. There has also been an implementation of an agent which modifies its dialogue strategy according to the engagement level of the user \cite{Yu2016}. We intend for ERICA to be used in a number of conversational scenarios such as interviewing, so the ability to detect the engagement level of the user is crucial to maintain a positive interaction.

We must first detect the relevant social signals which indicate engagement. There have been numerous works which detail these behaviors and apply them to agents and robots. They include facial expressions \cite{Castellano2009,Yu2017}, posture \cite{Sanghvi2011}, conversational phenomena \cite{Rich2010,Xu2013} as well as the four social signals which we investigate in this work. Typically the models to recognize social signals are trained using machine learning techniques.

Our goal is to extend the research by not only recognizing social signals but creating the engagement recognition model itself. We propose a hierarchical Bayesian model. Rather than arbitrarily choosing which social signals are needed, we conducted an experiment to determine which are relevant as judged by multiple third party observers of an interaction \cite{Inoue2016}. This is a similar approach to a recent work \cite{Oertel2015}, where the engagement model is directly learned from the results of the annotation. Our model also considers the latent character of the observer as an underlying variable to accommodate for differing personalities. In this work we describe the detection of the social signals provided as a result of our experiment.

\section{Data Collection}\label{data}
\begin{figure}
	\centering
		\includegraphics{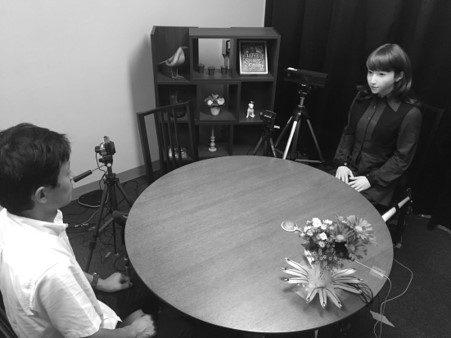}
	\caption{ERICA participating in a conversation with a user.}
		\label{ericaconvo}
\end{figure}

We created our own corpus to collect the input data for training. The corpus consists of 91 conversational sessions between a human subject and ERICA. Each session lasted between 5 and 20 minutes. An image of the scenario is shown in Figure \ref{ericaconvo}. Note the Kinect sensor to the right of ERICA and the vase on the table which is actually a microphone array.

In the conversation scenario, ERICA plays the role of a secretary for a laboratory in a university. The subjects arrive at the lab with the intention of talking to the professor, who is temporarily absent. ERICA informs the subject of this fact and while they are waiting begins a conversation. The topics of conversation varied and included the subject's hobbies and interests, life as a student, and their thoughts on android robots. ERICA asked and answered questions about herself according to rough guidelines set out before the experiment.

ERICA was controlled by a female operator in a remote, hidden location. During the sessions we used six different operators. All were professional voice actors. The subjects heard the operator's speech, which was also used to move ERICA's lips and mouth in a natural manner \cite{Sakai2015}. The operator could also control ERICA's head movements through the use of controllers.

We recorded several streams of data for the conversational sessions. These included multiple perspectives of video, audio channels of both ERICA and the subject, and motion capture data from a Kinect sensor which was located next to ERICA. The relevant data from the Kinect sensor is the yaw, roll and pitch values of the head captured at 30 Hz, relative to the Kinect sensor. We use several microphones to capture separate audio channels of ERICA and the user. Kinect and audio were synchronized.

To create a ground truth of the data, annotators watched videos of the sessions and marked the beginning and end points of each target behavior. We only consider behaviors which occurred during ERCIA's conversational turn. Due to the time taken to annotate the sessions, we do not use all of them when training the behavior models. Furthermore, the data used to train each of the behaviors are not necessarily consistent across all the models. This is because different annotators were responsible for different types of behavior and so worked independently on the sessions.

\section{Selection of Engagement-related Behaviors} \label{engagementsection}
Previously we conducted an experiment in which multiple third-party annotators watched the sessions of interaction and identified the behaviors which accompany a change in perceived engagement \cite{Inoue2016}. The outcome of this experiment was that nodding, laughter, verbal backchannels, and eye gaze were commonly chosen by the annotators as being indicative of engagement. The annotation of engagement was inconsistent among the third-party annotators, so we also modeled the character of the annotator as a latent variable. Our experiment also showed that facial expression was also important, but because this is more difficult to annotate, we omit it from our model.

We focus on engagement as being a function of the behaviors of the human in their role as a listener. In our system we have three main tasks. The first is to measure behaviors of the user during the system's speaking turn, which is the focus of this work. Secondly, we estimate their level of engagement based on the previously mentioned behaviors using a hierarchical Bayesian model. Thirdly, we choose the robot's dialogue strategy for her next turn based on the engagement level. This is to be considered in the near future.

The type of classification for each behavior is slightly different. Nodding is detected continuously as soon as new output from the Kinect sensor is received. Laughter and backchannel detection is detected whenever a new inter-pausal unit (IPU) is said by the user. In the live system, this IPU is an output of the Japanese automatic speech recognition (ASR) system Julius \cite{Lee01}. Eye gaze is detected per speaker turn.

The integrated system can recognize the engagement level during the system's speaking turn. The only required hardware is a Kinect sensor and a microphone array to implement this system, so we expect that it can be used for other conversational robots. In the next sections we will describe the behavioral models and their performances in detail.

\section{Nodding}
Nodding is often used as a backchannel by the listener, particularly in Japanese conversations \cite{Maynard1987,Hanzawa2012}. Nodding was also identified by annotators as being indicative of engagement. Several nodding detection algorithms have been implemented in previous works using machine learning methods such as hidden Markov models, support vector machines and hidden conditional random fields \cite{Fujie2004,Morency2005,Wang2006}. We used a long short-term memory (LSTM) network \cite{Hochreiter1997} for the head nodding model, which can be readily applied to gesture recognition \cite{Ordonez2016}.

Input to this model consists of a sequence of frames from the Kinect sensor. The LSTM was constructed using Tensorflow using a GPU on a desktop computer. From preliminary trials we found that using a sequence of 30 frames (approximately 300ms of data) was suitable for training. For each frame we calculated 7 features. The most basic of these were the instantaneous speeds of the yaw, roll and pitch of the head. We also included features related only to the head pitch over the previous 15 frames. These were the average speed, average velocity, acceleration and range. To classify a sequence as a nod, we observe whether the final frame in the sequence is within the range of a nod in the ground truth annotation.

We used an LSTM with 16 nodes, a learning rate of 0.001 and equally weighted samples. The mini-batch size was fixed at 32 samples. Every five epochs we measured the error on the validation set to determine early stopping. This epoch was then used as a parameter for evaluating the test set.

\begin{table}[h]
\begin{center}
\begin{tabular}{|l|r|r|}
\hline & \bf Median & \bf IQR \\ \hline
Listener turns per session & 38 & 8.5\\
Length of listener turns (s) & 4.89 & 7.03\\
Nods per session & 31 & 30.5\\
Length of nods (s) & 0.74 & 0.50\\
\hline
\end{tabular}
\end{center}
\caption{\label{corpus} Statistics of the 19 conversation sessions used for the nodding model. Distributions are highly skewed so we report the median and inter-quartile range (IQR).}
\end{table}

Due to the large amount of data and time to do the annotation, we used only 19 sessions in this work. Table \ref{corpus} displays general statistics about the subset of the corpus. We extracted 27,360 samples from the corpus, with 3,152 samples in the nodding class (approximately 11.5\% of the total). We performed 10-fold cross-validation across 19 sessions, with each fold consisting of two sessions for testing and two sessions for validation, except for one fold which contained only one session each for validation and testing.


We compared our model to a baseline model which always predicts the positive class, a support vector machine (SVM) with a radial basis kernel, a single hidden layer neural network and a deep neural network (DNN) with two hidden layers. We used the same features as the LSTM, but combined them in one input vector with size 210 (30 frames x 7 features). The neural network and DNN used the same learning rate and mini-batch size as the LSTM. The number of nodes of the hidden layers in the neural network and DNN were 128.

We performed a frame-wise evaluation of the models. The performance across all test sets for the positive class is shown in Table \ref{nodresults}.

\begin{table}[h]
\begin{center}
\begin{tabular}{|l|l|l|l|l|}
\hline \bf Nodding Model & \bf Prec. & \bf Rec. & \bf F1 & \bf Acc. \\ \hline
Baseline & 0.115 & 1.000 & 0.206 & 0.796\\
Neural network & 0.482 & 0.411 & 0.444 & 0.881\\
SVM (radial basis) & 0.424 & 0.542 & 0.475 & 0.863\\
DNN & 0.511 & 0.516 & 0.514 & 0.887\\
LSTM & 0.566 & 0.589 & 0.577 & 0.901\\
\hline
\end{tabular}
\end{center}
\caption{\label{nodresults} Performance of frame-wise nodding detection models.}
\end{table}

For frame-wise detection, the LSTM approach outperforms the other models over all metrics. We also evaluated the models by detecting nodding sequences rather than individual frames. We set a minimum length of a nodding sequence $l$, and only label a nod if it is continuously detected as such for longer than $l$ milliseconds, to reject short, isolated sequences of nods. If a detected nod sequence from the model overlaps with a ground truth nod sequence, then we label it as correct. We evaluated various values of $l$ and found that setting this value to 300ms was optimal for all the models. Results are shown in Table \ref{nodresultsevent}.

\begin{table}[h]
\begin{center}
\begin{tabular}{|l|l|l|l|}
\hline \bf Nodding Model & \bf Prec. & \bf Rec. & \bf F1\\ \hline
Neural network & 0.551 & 0.581 & 0.566 \\
SVM (radial basis) & 0.550 & 0.638 & 0.591 \\
DNN & 0.553 & 0.687 & 0.613 \\
LSTM & 0.608 & 0.763 & 0.677 \\
\hline
\end{tabular}
\end{center}
\caption{\label{nodresultsevent} Performance of event-wise nodding detection models.}
\end{table}

We again find that the LSTM is the best performing model and that event-wise detection is better than frame-wise detection.

\section{Laughter}
Some works have used smiling to detect engagement \cite{Castellano2009,Yu2017}. Smiling in general displays a positive feeling toward the agent and so it is natural that this would indicate engagement. However, reliably recognizing smiles would require us to integrate a camera into our system. As a proxy, we hypothesize that laughter also indicates engagement. Furthermore, this was also confirmed by the third party annotators in our experiment. For these reasons we opt to build a system that can detect laughter, making use of the microphone array that is already used by the system. Automatic laughter detection is a well-researched topic \cite{Cosentino2016}.

We constructed a binary classifier to detect whether or not a given IPU contains laughter. For each IPU, we extracted prosodic and linguistic features. We use the duration and the voiced-unvoiced intensity ratio of the IPU plus six features related to its pitch and the intensity (the mean, median, slope, minimum, maximum and range), giving a total of 14 prosodic features. For linguistic features, we include the classifications of the previous 5 IPUs, a normalized word representation of previous utterances provided by Word2Vec \cite{Mikolov2013}, and part of speech tags provided by JUMAN++ \cite{Morita2015}. A total of 55 linguistic features are used. 

We classified an IPU as containing laughter whether it occurred in isolation or as part of an utterance. We used 46 sessions of the corpus to test the laughter detection system. Our corpus contained 9,320 IPUs, with the percentage of laughter samples being 6.3\%. Cross-validation was performed leaving one session out for each fold. We tested various models and found that a two-layer DNN had the best performance.

Results of the laughter detection model are shown in Table \ref{laughresults}. We compare two type of models to the baseline. The first uses only prosodic features, while the second adds linguistic information based on the transcription of the corpus.

\begin{table}[h]
\begin{center}
\begin{tabular}{|l|l|l|l|l|}
\hline \bf Model & \bf Prec. & \bf Rec. & \bf F1 & \bf Acc. \\ \hline
Baseline & 0.063 & 1.000 & 0.119 & 0.882 \\
Prosody only & 0.343 & 0.222 & 0.269 &  0.927 \\
Prosody + linguistic & 0.587 & 0.455 & 0.513 & 0.947 \\
\hline
\end{tabular}
\end{center}
\caption{\label{laughresults} Performance of laughter detection models.}
\end{table}

The addition of linguistic features improves the performance of the model, because of transcriptions such as ``hahaha'' which explicitly represent laughter. However, we cannot guarantee that an ASR system would successfully generate the same output as the transcription. In this case, a prosodic-only model provides an alternative solution. We acknowledge that the inclusion of other spectral features would considerably improve the model.

\section{Verbal backchannels}
Backchannels are short responses to the speaker during a conversation, such as \textit{mm} and \textit{uh-huh} in English or \textit{un} and \textit{ee} in Japanese, where they are termed aizuchi. These occur more frequently in Japanese than in English and are often accompanied by head movements \cite{Ike2010}. Backchannels have been addressed in several works as being an indicator of engagement \cite{Rich2010,Oertel2015} and confirmed as being so by annotators in our experiment. 

Our goal is to create a model which can effectively recognize backchannels during the robot's speaking turn. Although our main task is to use the results of this recognition for our engagement model, it is also useful to classify backchannels because the system should distinguish them from situations where the user is actually starting their speaking turn or trying to barge-in during her speech. 

We again use IPUs as an input and construct a binary classifier to determine if an IPU is a backchannel or not. We used the same sessions for training as the laughter detection model. 26.2\% of samples were classified as verbal backchannels. We also used the same features as the laughter detection model and found that the best model was a random forest with 56 estimators. Table \ref{bcresults} shows the results of backchannel detection. As with laughter detection, we compared the model with and without linguistic features. 

\begin{table}[h]
\begin{center}
\begin{tabular}{|l|l|l|l|l|}
\hline \bf Model & \bf Prec. & \bf Rec. & \bf F1 & \bf Acc. \\ \hline
Baseline & 0.262 & 1.000 & 0.415 & 0.613 \\
Prosody only & 0.818 & 0.745 & 0.780 & 0.887 \\
Prosody + linguistic & 0.909 & 0.926 & 0.918 & 0.955\\
\hline
\end{tabular}
\end{center}
\caption{\label{bcresults} Performance of backchannel detection models.}
\end{table}

We again see that linguistic features improve the performance of the model. This is expected as backchannels have several common lexical forms. However, even using only prosodic features the model still has a reasonable F1 score so we do not need to rely on a correct ASR output.

\section{Eye gaze}
Eye gaze behavior has been identified in previous research as indicative of engagement \cite{Nakano2010,Rich2010} and the annotators in our experiment confirmed this. We consider that the most important aspect of the user's eye gaze is that they are looking at ERICA while she is speaking.

Although eye-gaze models using computer vision techniques have been developed in previous work \cite{Sewell2010,Zhang2015}, we opted to use a simpler geometry-based method with the Kinect sensor. We define the 3-d world co-ordinates of ERICA and the Kinect sensor's position, and can receive the vector of the head orientation provided by the Kinect sensor. We first transform this orientation into world co-ordinates so that we have a vector whose origin and direction are the user's head and head orientation. We then use collision detection to check for intersections between the vector and a 30cm sphere around ERICA's head. This is to accommodate for the measurement of head orientation rather than actual eye gaze. If there is an intersection, we label it as looking at ERICA.

From our experiment we found that the highest inter-annotator agreement of engagement (Spearman's correlation coefficient of $0.375$) was when the subject gazed at ERICA continuously for 10 seconds during her speaking turn. Therefore, we use this rule as a basis for the eye gaze input. Turns of less than 10 seconds long were classified as negative according to the eye gaze model. We manually annotated a ground truth of eye gaze through visual observation then labelled the turn according to the 10 second rule. We then generated labels according the output of our eye gaze model. The number of turns which were labeled as positive (the user gazed at ERICA for at least 10 seconds continuously) was 17.1\% of the total. The model was tested using 20 sessions of data and results are shown in Table \ref{gazeresults}.

\begin{table}[h]
\begin{center}
\begin{tabular}{|l|l|l|l|l|}
\hline \bf Model & \bf Prec. & \bf Rec. & \bf F1 & \bf Acc. \\ \hline
Baseline & 0.171 & 1.000 & 0.292 & 0.716 \\
Gaze model & 0.504 & 0.580 & 0.539 & 0.847 \\
\hline
\end{tabular}
\end{center}
\caption{\label{gazeresults} Performance of eye gaze detection model.}
\end{table}

We find that the model works reasonably well for classifying continuous gaze behavior. However we could only estimate the positions of ERICA's head and the Kinect sensor in our corpus by observing the video. In the live system we calculate these values exactly, so we expect that the performance of the model will be improved from this result.

\section{Engagement Recognition}
\begin{figure}[h]
	\centering
		\includegraphics{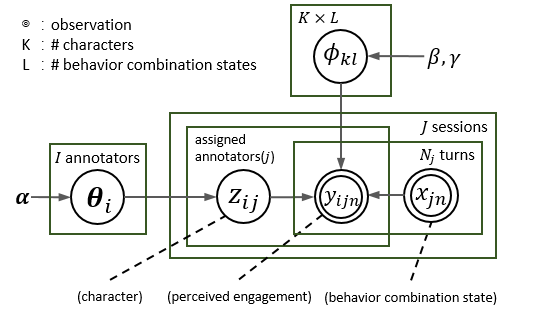}
	\caption{Graphical model of engagement recognizer}
		\label{model}
\end{figure}

We have described four different social signals for and their recognition models. The performance of the models is varied, but our goal is not to produce state-of-the-art individual models, but to assess whether they can be used in conjunction with our engagement recognition model. This is a hierarchical Bayesian binary classifier, predicting if a listener is engaged or not during the system's speaking turn. The graphical model is shown in Figure \ref{model}.

For our evaluation we selected 20 sessions from our corpus which were a different subset than those used to train the individual social signal models. We recruited 12 third-party observers to watch video of these sessions and annotate time-points where they perceived the engagement of the subject was changed to high. We defined engagement for the annotators as ``How much the subject is interested in and willing to continue the current dialogue with ERICA''. The annotator variable is represented by $i$ in our graphical model.

The input to the model, $x_t$ is a state which is defined as the observed combination of social signals during a turn. Each social signal can be classified as a binary value (either detected or not detected), giving 16 possible combinations. When training the model we know this combination, but in a live system we marginalize over each behavior combination using its prior probability.

We also define a variable $k$, which represents a specific character type of an annotator. For example, laughter may be influential for one character type but not so important for another. Therefore, annotators with different characters would perceive laughter differently in terms of perceived engagement. We created a distribution of characters for each annotator and found that they could be grouped by similar character. From our experiments we found that including a character variable improved the model's performance and that the best model had three different character types ($K=3$).

For brevity, we omit details of most of the calculations used in our model. The posterior probability of perceived engagement is 

\begin{equation}
p(y_{it}|x_t,i,\tilde{\Theta},\tilde{\Phi})=\sum_{k=1}^{K}{\tilde{\theta}_{ik}\tilde{\phi}_{kx_t}}
\end{equation}

where $x_t$ is the observed behavior combination for a system turn, $i$ is the index of the annotator, and $\Theta$ and $\Phi$ are the learned parameters of the latent character model. $\tilde{\theta}_{ik}$ is the probability that an annotator $i$ has a character $k$, and $\tilde{\phi}_{kx_t}$ is the probability that the behavior combination $x_t$ is perceived as engaged by $k$. 

We compared the performance of our engagement model under two conditions - using manually annotated data as inputs and using the results of our social signal detection models as inputs. Engagement is classified per the robot's speaking turn.  We also analyze the model according to whether it uses contextual information. This means the result of engagement in the system's previous turn is used as a feature for classifying engagement in her current turn. We use the area under the precision-recall curve (AUC) as a performance measure. The results are shown in Table \ref{engagementresults}.

\begin{table}[h]
\begin{center}
\begin{tabular}{|l|r|r|}
\hline \bf Labeling method & \bf No context & \bf Context \\ \hline
Manual annotation & 0.650 & 0.669  \\
Detection system & 0.615 & 0.620 \\
\hline
\end{tabular}
\end{center}
\caption{\label{engagementresults} AUC scores for the engagement model using manual annotation and our social signal detection models.}
\end{table}

From our results we see a drop in performance of the engagement model when using social signal detection compared to manual annotation, but is not drastic. We also see that adding contextual information provides an improvement in performance. We can also show that even though the individual models do not all have high recognition ability, their combination is adequate for engagement recognition.

\section{Discussion}
We find that the individual social signal models have varying levels of performance. Laughter detection is quite poor, while backchannel detection is considerably better. To improve the performance of our recognition systems, we could add other modalities as in other works \cite{Morency2005}. In particular, the co-gesture of verbal backchannels and nodding could help to improve both systems. Similarly, visual recognition of laughing would improve the results of laughter detection. We are trying to improve the performance of the individual models, including using spectral features for better laughter and backchannel detection.

Although we have evaluated the model using our corpus, we expect that laughter and backchannel detection performance in the live system will be slightly degraded because these signals are mixed in with ERICA's speech during her turn. We limit this by using the microphone array to ignore ERICA's voice, but cannot guarantee that user speech will be clean. Other previous works tend to focus on non-verbal social signals, but from our third party annotation experiment, verbal signals are necessary.

Our data was collected in a one-to-one conversational setting, but the models are not restricted to this specific environment. Nodding, laughter and backchannel detection are independent of both environment and user. The eye gaze model needs to be calibrated to accommodate the position of Kinect and ERICA. We have successfully implemented all our models in a separate environment, with different placements of Kinect and ERICA. We propose that the models can function in a varied number of conversational settings, including multi-party dialogue.

Our next step is to use the results of engagement recognition to modify the dialogue policy of the system. We consider that the engagement of the user has an influence on turn-taking behavior or changing the topic of conversation. However, we can also consider that the flow of dialogue may be completely modified. One scenario we are considering is ERICA giving a technical explanation. By recognizing if the user is engaged, the robot may use simpler terminology to make her talk more understandable. We intend to formulate such scenarios where conversational engagement recognition is necessary and then conduct user experiments to confirm the effectiveness of our system.

\section{Conclusion}
This paper described models for detecting nodding, laughter, verbal backchannels and eye gaze, which will be used by an engagement recognizer during conversation with a robot. The robot we use in this work is the android ERICA. We selected these social signals based on a previous experiment where third party observers annotated changes in engagement based on behaviors. The inputs are a Kinect sensor and a microphone array. Although the performance of our models are varied, their combination is effective. We observe a slight degradation in performance for our engagement recognition model when using the outputs of social signal detection compared to annotated values. We have integrated these models into ERICA's system architecture and intend to make her an engagement-aware conversational robot.

\section{Acknowledgements}
This work was supported by JST ERATO Ishiguro Symbiotic Human-Robot Interaction program (Grant Number JPMJER1401), Japan.

\bibliography{NCHRC}
\bibliographystyle{aaai}
\end{document}